\documentclass[apjl]{emulateapj}
\usepackage[latin2]{inputenc}
\usepackage{amsmath}
\usepackage{graphicx}
\usepackage{url}
\usepackage{natbib}

\begin{document}
\title{Update on the Nature of Virgo Overdensity}
\author{Ana Bonaca\altaffilmark{1, 2}, Mario Juri\' c\altaffilmark{3, 4}, \v Zeljko Ivezi\'c\altaffilmark{5}, Dmitry Bizyaev\altaffilmark{6}, Howard Brewington\altaffilmark{6}, Elena  Malanushenko\altaffilmark{6}, Viktor Malanushenko\altaffilmark{6}, Daniel Oravetz\altaffilmark{6}, Kaike Pan\altaffilmark{6}, Alaina Shelden\altaffilmark{6}, Audrey Simmons\altaffilmark{6}, Stephanie Snedden\altaffilmark{6}} 
\altaffiltext{1}{Department of Astronomy, Yale University, New Haven, CT 06511; {ana.bonaca@yale.edu}}
\altaffiltext{2}{Department of Physics, Faculty of Science, University of Zagreb, Croatia}
\altaffiltext{3}{Institute for Theory and Computation, Harvard-Smithsonian Center for Astrophysics, Cambridge, MA 02138; {mjuric@cfa.harvard.edu}}
\altaffiltext{4}{Hubble Fellow}
\altaffiltext{5}{Department of Astronomy, University of Washington, Seattle, WA 98195;}
\altaffiltext{6}{Apache Point Observatory, P.O. Box 59, Sunspot, NM 88349;}

\begin{abstract}

We use the Eighth Data Release of Sloan Digital
Sky Survey (SDSS DR8) catalog with its additional sky coverage of 
the southern Galactic hemisphere, to measure the extent and study 
the nature of the Virgo Overdensity \citep[VOD;][]{J08}. The data show that the VOD 
extends over no less than 2000 deg$^2$, with its true extent likely closer to 3000 deg$^2$.
We test whether the VOD can be attributed to a tilt in the stellar halo 
ellipsoid with respect to the plane of the Galactic disk and find that the
observed symmetry of the north-south Galactic hemisphere star counts
excludes this possibility.  We argue that the Virgo Overdensity, in spite of
its wide area and cloud-like appearance, is still best explained by a minor merger.
Its appearance and position is qualitatively similar to a near peri-galacticon
merger event and, assuming that the VOD and the Virgo Stellar Stream
share the same progenitor, consistent with the VSS orbit determined
by \cite{dinescu2009}.

\end{abstract}
\maketitle

\section{Introduction}

A growing number of observational campaigns over the last decade showed the
richness of substructure in the Milky Way halo.  The most vivid discoveries
of ongoing accretion onto the Galaxy include the Sagittarius dwarf
and its tidal streams \citep[][among others]{I00,yanny2000,vivas2001,majewski2003}, 
the Monoceros stellar stream \citep{newberg2002,rp2003}, as well as
numerous smaller streams and dwarf galaxies abundant in the so-called ``field of
streams" \citep{belokurov2006}.  Evidence of ongoing mergers has also been
detected in the halo of M31 \citep{ibata2001, ferguson2002, morrison2003,
zucker2004, kalirai2006, fardal2007}, providing further support for the
importance of hierarchical merging in galaxy formation.  

Even if one is
ultimately interested in the processes of galaxy formation in general, the
study and understanding of these processes in the Milky Way remains
advantageous in many aspects.  For example, the proximity of stars and
substructures in the Galaxy allows data collection at high resolution,
which can be directly tested against predictions from simulations.

In this work, we concentrate on the Virgo Overdensity \citep[VOD;][]{J08}, a 
cloud-like overdensity of stars spanning distances between $10-20$ kpc.  Parts of the
VOD were initially identified as an overdensity of RR Lyrae stars confined
to a small region in the Quasar Equatorial Survey Team survey
\citep{vivas2001}, followed by an identification as an overdensity of main
sequence turnoff stars in the same direction \citep{newberg2002}.  However,
it was not until the advent of wide-area surveys such as the Sloan Digital
Sky Survey \citep[SDSS;][]{york2000} that the size of the VOD feature was fully
recognized \citep{J08}.  In particular, using a large photometric sample of
main sequence stars, \citet{J08} have shown that the VOD
extends over at least 1000 deg$^2$, spans heliocentric distances between $6
< D < 20$~kpc, and also shows as an overdensity of M giants in the Two
Micron All Sky Survey \citep[2MASS;][]{skrutskie2006}.  Based on their
$u-g$ band color, they further argued for low metallicity of its constituent
stars. This is consistent with the spectroscopic determination by \citet{duffau2006}
([Fe/H]=$-1.9$), who identified a velocity peak of RR Lyrae
stars in the same direction, naming it the ``Virgo Stellar
Stream" (VSS)\footnote{In this paper, we use the term ``Virgo
overdensity" to denote both the excess stellar number density, as well as
the kinematic peaks observed in the general directon of Virgo at distances
$\sim 5-25$ kpc as it is unclear at present whether the two are truly distinct.}. Subsequent
kinematic studies showed that there may be individual filaments within the
VOD/VSS region \citep{vivas2008}, likely to belong to more than one stream. 
The current knowledge of VOD/VSS properties is summarized in Table~\ref{overview}.

Despite these extensive studies, there is still no definitive answer on the
origin of the VOD/VSS nor of its extent.  Since the stars are metal poor, but cover
a wide range of metallicities, both \citet{duffau2006} and \citet{J08} have
argued it to be the debris of a tidally disrupted dwarf spheroidal galaxy. 
Supporting this hypothesis, \citet{dinescu2009} showed that orbit of an RR
Lyrae star associated with the VSS, derived from proper motion and radial
velocity observations, is consistent with a merging scenario in its early
phase.  An alternative, which would explain the unusually large angular size
of the VOD, is that the overdensity may not
be due to a merger but a misinterpretation of a more complex large-scale 
structure of the Galactic stellar halo. However, given the limited extent of
wide-field data available at the time, this hypothesis could not be tested
conclusively.

Recently, the Eighth Data Release (DR8) of the Sloan Digital Sky Survey has
become public \citep{dr8}.  Compared to previous releases, DR8 adds a
significant new area ($\sim 2000$ deg$^2$) in the southern Galactic
hemisphere, as well as filling in a few ``holes" in the northern footprint. 
Importantly, the added area allows us to conduct {\em symmetry studies},
model-free comparisons of stellar number densities along directions where we
expect them to be the same if the halo conforms to a given shape (spherical,
ellipsoidal, etc.).

In this paper, we use this newly available dataset to remeasure the size of
VOD, better understand its large-scale structure, and attempt to discern between the
two proposed options for its origin.  We begin this analysis in
Section~\ref{data} with an overview of the used stellar sample, details on the
photometric parallax, and construction of stellar number density maps. 
Section~\ref{res} brings analysis of density maps, with special attention
given to the extent of the overdensity and how it compares to surrounding
area without overdensities. In the final Section we discuss the implications of
presented results on the nature of Virgo.

\begin{table}
\caption{Overview of Virgo Overdensity / Stellar Stream properties}
\label{overview}
\begin{center}
\begin{tabular}{p{2cm} p{2.8cm} p{2.8cm}}
\hline\hline
\sc Quantity & \sc Value & \sc Reference \\
\hline\hline
Angular size (deg$^2$) & $>106$\footnotemark[1] & \citet{duffau2006} \\
 & $>1000$ & \citet{J08} \\
 & $\sim760$\footnotemark[1] & \citet{prior2009} \\
 & $>2000$ & this work \\
\hline
Surface brightness (mag arcsec$^{-2}$) & 32.5 & \citet{J08} \\
\hline
Distance (kpc) & 20 & \citet{newberg2002} \\
 & 19 & \citet{vivas2003} \\
 & 6 to 20 & \citet{J08} \\
 & 19\footnotemark[1] & \citet{prior2009} \\
 & 15 to 30 & \citet{brink2010} \\
\hline
Metallicity [Fe/H] & $-1.86\pm0.40$\footnotemark[2] & \citet{duffau2006} \\
 & $-1.5$\footnotemark[3] & \citet{J08} \\
 & $-2.0\pm0.1$ (internal)\par $\pm0.5$ (systematic)\footnotemark[3] & \citet{an2009} \\
\hline
Radial velocity (km s$^{-1}$) & $99.8\pm17.3$\footnotemark[1] & \citet{duffau2006} \\
 & $130\pm10$ & \citet{newberg2007} \\
 & $127\pm10$\footnotemark[1] & \citet{prior2009} \\
 & $134.4\pm14.0$\footnotemark[1] & \citet{dinescu2009} \\
\hline
Proper motion (mas yr$^{-1}$) & $\mu_\alpha\cos{\delta}=-3.50\pm0.85$, $\mu_\delta=2.33\pm0.85$ & \citet{dinescu2009} \\
\hline
Origin & Sagittarius dSph & \citet{md2007}\\
 & dwarf galaxy & \citet{J08, dinescu2009}; this work\\
\hline\hline
\end{tabular}
\end{center}
 \footnotetext[1]{Quantity related to the VSS.}
 \footnotetext[2]{Spectroscopically derived quantity.}
 \footnotetext[3]{Photometrically derived quantity.}
\end{table}

\section{Data and Methodology}
\label{data}

In this section we describe characteristics of the SDSS imaging survey and the
subset of its stellar sample used in this work. We discuss how the distance to
each star was calculated using the photometric parallax method, paying
special attention to metallicity effects. In conclusion, we show how the
stellar density maps were created.

\subsection{SDSS DR8 Imaging Survey}

The Sloan Digital Sky Survey III provides an unabridged view of
the night sky.  Its total imaging footprint covers 14,555 deg$^2$, a third
of the celestial sphere, including $\sim$5,200 deg$^2$ of imaging on the
Southern Galactic Hemisphere available in Data Release 8 \citep{dr8}.  All
of the planned imaging is now complete, but spectra will continue to be taken
until the project ends in 2014. These will further enlarge the
existing library which already contains spectra of 520,000 stars,
860,000 galaxies and 120,000 quasars. For a detailed overview of ongoing spectroscopic projects see \citet{eisenstein2011}.

The SDSS photometric component collects data in five optical bands: $u$, $g$, $r$, $i$
and $z$ measured in $AB_\nu$ magnitude system \citep{gunn1998, fukugita1996, gunn2006},
with 95\% completeness levels at magnitudes 22.1, 22.4, 22.1, 21.2, and 20.3 
respectively.  The latest data release has very accurate purely internal photometric
calibration \citep[so-called, {\em \" ubercalibration}]{padmanabhan2008}, with
the same standard-star derived zeropoints used to absolutely calibrate the 
previous releases. Furthermore, all the imaging data was reprocessed with
the new photometric pipelines, featuring enhanced sky-subtraction algorithm.
Unfortunately, due to a mistake in data processing, the absolute astrometry of DR8 is less
accurate than the previous releases (M. Blanton; private communication), but
this is of no consequence to the work presented in this paper.

Importantly, the morphological star-galaxy separation algorithm is well 
understood and has not changed since DR2. It classifies objects as 
`GALAXY' if:
\begin{equation}
	\mathrm{psfMag} - \mathrm{cmodelMag} > 0.145.
\end{equation}
This admittedly simple criterion has been shown to work well for selecting
clean samples of stars to $r \sim 21.5$ \citep{J08}.

\subsection{The Photometric Parallax Relation and Iterative Determination of
Distances}

More than 95\% of stars detected by the SDSS are on the main sequence
and of similar age, therefore residing on a fairly constrained,
one-dimensional, stellar locus. Provided we have a calibrated
color-luminosity (or, {\em photometric parallax}) relation, this fact
allows us to estimate both their absolute magnitudes and distances from
multi-band photometry.

Once the absolute magnitude of a star is known, its distance is easily calculated using:
\begin{equation}
D (\text{pc}) = 10^{(r-M_r)/5+1}
\label{distance}
\end{equation}
which then leads to its position in Galactocentric Cartesian coordinates:
\begin{eqnarray}
X &=& R_\odot - D\cos{l}\cos{b} \nonumber \\
Y &=&  -D\sin{l}\cos{b}\\
Z &=&  D\sin(b)\nonumber
\label{position}
\end{eqnarray}
where $R_\odot$ is distance from Sun to the Galactic center and $l$ and $b$
are Galactic longitude and latitude, respectively.

There have been numerous efforts through studies of stellar systems with
known distances (eg. nearby stars, globular clusters) yielding a number of
proposed photometric parallax relations \citep[eg.][]{hawley2002, williams2002, west2005, bilir2006}. In this
paper, we use the relation derived by \citet{I08}:
\begin{equation}
M_r(g-i, [\text{Fe/H}]) = M_r^0(g-i)+\Delta M_r([\text{Fe/H}]) \\
\label{parallax} 
\end{equation}
where the terms $M_r^0(g-i)$ and $\Delta M_r([\text{Fe/H}])$ have been
determined to be:
\begin{eqnarray}
&M&_r^0(g-i) = -0.56 + 14.32(g-i) - 12.97(g-i)^2 \nonumber \\ 
&&+ 6.127(g-i)^3 - 1.267(g-i)^4 + 0.0967(g-i)^5 \label{parallax2} \\ 
&\Delta&M_r([\text{Fe/H}]) = -1.11([\text{Fe/H}]) - 0.18([\text{Fe/H}])^2 \nonumber
\end{eqnarray}
This relation was calibrated from SDSS observations of 11 star clusters in
metallicity range from $+0.12$ to $-2.50$.  It builds on
the previous work of \citet{J08}, and is in good agreement with other
proposed relations in the literature. It is expected to be accurate to $\sim
10- 15$\% \citep{I08}, and the fact that it was directly calibrated on SDSS
photometry makes it especially apropriate for use here.

While more accurate, the \citet{I08} relation is at a disadvantage compared
to, for example, the \citet{J08} due its need for the $u$-band
photometry to determine metallicity.  Given the shallower depth of
the SDSS $u$ band observations, this would effectively limit our
explorations to $\sim8$ kpc, heliocentric. We therefore chose not to determine metallicity from the photometry, but instead use the
metallicity prior given by \citet{I08} models to compute the
$\Delta M_r([\text{Fe/H}])$ needed in Equation~\ref{parallax2}. Since
\citet{I08} give the metallicity as a function of position in the Galaxy
(that, in turn, depends on the absolute magnitude), this makes our problem
an implicit one.

We iteratively solve for $D$ and $M_r$ starting with an initial guess for
metallicity of [Fe/H$]=-0.5$.  We then compute $D$ and $M_r$ from the
observed $g, i$.  Next, the expected metallicity at that position in the
Galaxy is drawn from distributions given by \citet[Equations~18--20]{I08}, and the process is
repeated until convergence is obtained.  Note that since \citet{I08} give their
metallicity distributions separately for the disk and the halo, we randomly
assign a star to the disk or halo component, with the weight given by their
local contributions.  In practice, this detail is of small importance for
stars at approximate distance of the VOD, as nearly all belong to
the halo.

We have compared the stellar number density dependence on the position in the
Galaxy as obtained with the iterative approach, to results of \citet{I08} as
well as the result obtained using \citet{J08} photometric parallax relation.
We found them to be in agreement to within $\sim$10\% in the thin and
thick disk regions, and discrepant with respect to the normalization of
the halo component. In particular, while the overall halo density
profile given by \citet{J08} is correct, we have found their local
halo-to-disk normalization to be over-estimated by approximately a factor of
three.

For Galactic models used in subsequent sections, we have lowered the
halo-to-thin disk normalization parameter $f_H$ from the \citet{J08} value
to $f_H=1.6\times10^{-3}$. The difference is not surprising in light of
a relatively large error bar ($50\%$) that \citet{J08} have attached to
their measurement of $f_H$. We also note that the value used here
is more consistent with those traditionally used. A formal fit for the $f_H$ is beyond
the scope of this work and will be discussed in a subsequent paper (Juri\' c
et al.; in prep). In here, we will refer to it as the {\em galfast} model.

\subsection{Sample Selection and Maps of Stellar Number Density}

To be included in our sample of main sequence stars, we require the objects to be morphologically classified as `STAR' by the
SDSS pipeline, be within $\sim$0.32 mag of the stellar locus as defined by
Equation~4 of \citet{J08}, and have colors $g-r>0$.  The last cut
removes extremely blue point sources, such as BHB stars or blue stragglers. We further require the objects to have $r, i < 21.5$, to restrict ourselves
to the range where SDSS' morphological star-galaxy separation works well
\citep{J08}.  The apparent magnitudes of the resulting sample, containing
$\sim86$ million objects, are then corrected for interstellar extinction using the SFD
maps \citep{sfd}.  Finally, we apply the iterative
distance/absolute-magnitude determination procedure to all objects (stars)
of this sample.

We proceed to divide up the sample in 6 absolute magnitude bins (spanning
$3.5 < M_r < 9.5$), and 9 shells of heliocentric distance (spanning $2 < D/\text{kpc} <
20$).  The bin sizes in absolute magnitude and distance are 1 mag and 2 kpc,
respectively. Results shown here contain only stars from the brightest 
absolute magnitude bin ($3.5 < M_r < 4.5$), which probe deepest into the stellar halo.

These data are then binned spatially, in Lambert equal area projection, and plotted in
hemisphere plots.  The projection poles were set to the north and south
Galactic pole for examination of northern and southern SDSS data,
respectively.  The Lambert projection was chosen as it allows for
straightforward comparison of sizes of observed structures on different
parts of the sky.  The map pixel scale of $dx=1.5^\circ$ was chosen to keep
most of the pixels well populated, thus reducing Poisson noise. At a central-distance bin of 9 kpc $\sim$70 \% of
pixels contain at least $\sim$30 stars, while at the largest distance of 19 kpc this fraction is $\sim$50 \%.

Not all pixels on the map have been fully covered by SDSS observations. This
is particularly true for pixels close to the edge of the footprint, and
needs to be accounted for. For consistent comparison of data and model predictions, we compute the
pixel fraction covered by the SDSS by subdividing each pixel with a 100x100 grid,
and counting the number of so defined subpixels that contain at least one
star. The fraction of pixel area covered by the SDSS is then approximated
by the ratio of counted subpixels, to the total number of subpixels. Model
predictions (discussed below) have been multiplied by this fraction.

Examples of stellar density maps thus obtained are given in Figures~\ref{denn} 
(north Galactic hemisphere) and \ref{dens} (south Galactic
hemisphere).  The panels in the left column show SDSS DR8 stellar counts, while
{\em galfast} model predictions for the same population of stars are given
in the center.  Model predictions give the expected number of stars in the volume
determined by pixel and distance bin size.  Finally, a quantitative
comparison of the data and model, given by their difference normalized to the
model, is shown on residual plots in the right column.

\section{Results}
\label{res}

In this section we discuss the density maps constructed as described above,
with emphasis on the structure of Virgo Overdensity.  Previous
studies found the VOD to peak at $\sim$10 kpc. Here we
present and discuss density distribution maps at two distances: $5\pm1$ kpc where we
do not see signature of the VOD, and $11\pm1$ kpc
where it is very prominent.

\subsection{Analysis of star counts maps}

\begin{figure*}
\begin{center}
\includegraphics[scale=0.9]{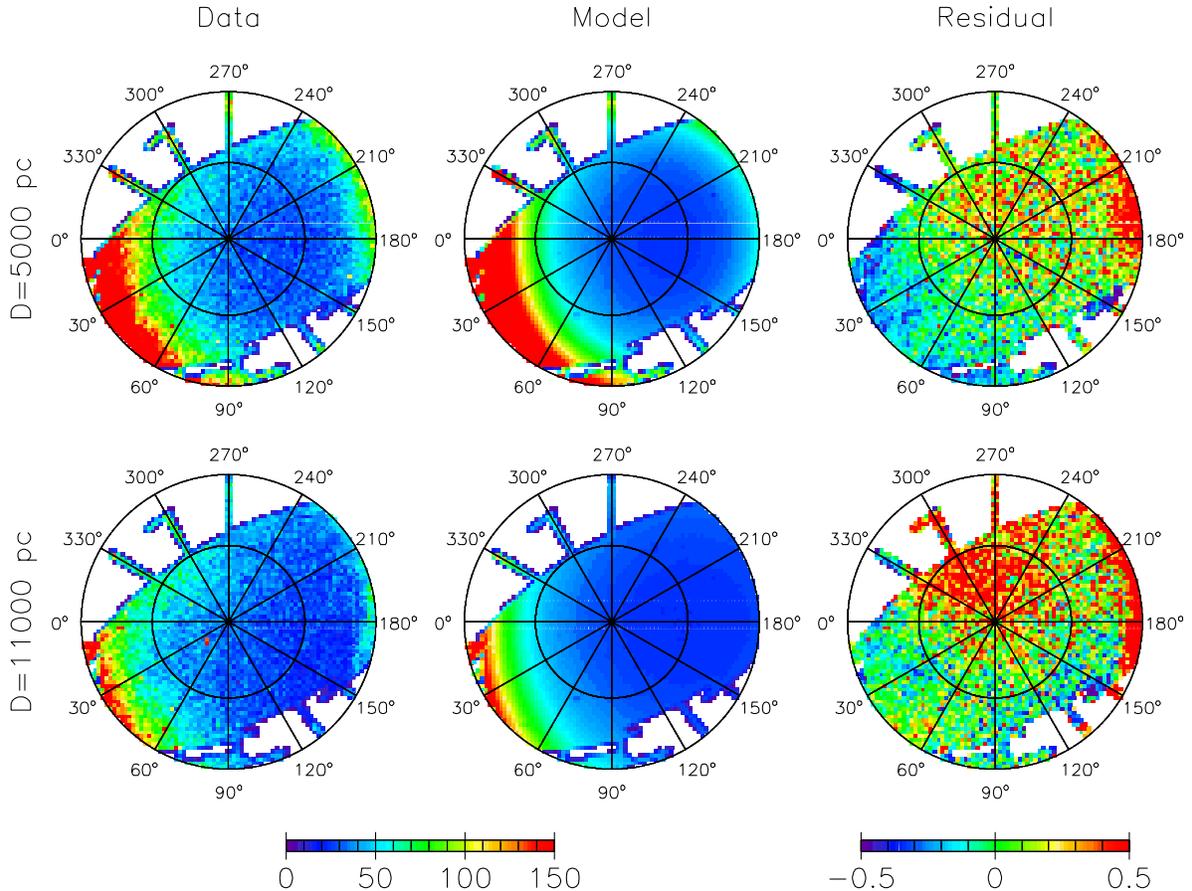}
\vspace{0.5cm}
\caption{Density distribution of F stars in 2 kpc wide shells centered on 5
and 11 kpc, mapped in Lambert equal area projection with circles
representing constant Galactic latitude and lines constant Galactic
longitude.  Disk area is excluded by plotting only latitudes higher than
30$^\circ$. Left panels show SDSS DR8 imaging data, with blank areas
corresponding to pixels with no data, best-fitting models updated from
\cite{J08} for the same area, corrected for SDSS sky coverage, are on the
middle panels, while the data - model residuals (normalized to the model)
are on the right panels.  At the 5 kpc distance shell, data and model are in
very good agreement (apart from the disk region, which was not modeled as carefully), 
but the 11 kpc shell clearly shows Virgo overdensity extending
over 2000 deg$^2$ in general direction of $l\sim300^\circ$, $b\sim75^\circ$. The Monoceros stream is visible at both distance shells as
overdensity on low Galactic latitudes for $150^\circ<l<230^\circ$.}
\label{denn}
\end{center}
\end{figure*}

As seen in Figure~\ref{denn}, the 5 kpc density map is well matched by the
model prediction, resulting in mostly uniform residual map centered around
zero.  The density map itself reveals the bulk structure of Milky
Way and our position inside it. The greatest density is observed at low
Galactic latitudes ($b\lesssim30^\circ$) where our distance shell dips
into the Galactic disk.  In order to reduce the clutter and obtain a better view of
the halo stars, the plots show only the region $b>30^\circ$.  In
comparison to the model predictions, we note two features on the map of
residuals.  First, the slight underdensity in the region heavily populated
by disk stars ($0^\circ<l<30^\circ$, $b<45^\circ$) due to model parameters
tuned to best reproduce the halo, and second the Monoceros stream viewed as
an overdensity on the opposite side of the map ($90^\circ<l<230^\circ$,
$b\lesssim40^\circ$).

The situation is radically different at 11 kpc. Even though the general
features seen in the 5 kpc plots are also present at 11 kpc, the residuals are
dominated by a large overdensity at $l\sim300^\circ$,
$b\sim75^\circ$ -- the Virgo Overdensity.  Probability of a random
fluctuation expanding over such an area is statistically insignificant.  

\begin{figure*}
\begin{center}
\includegraphics[scale=0.9]{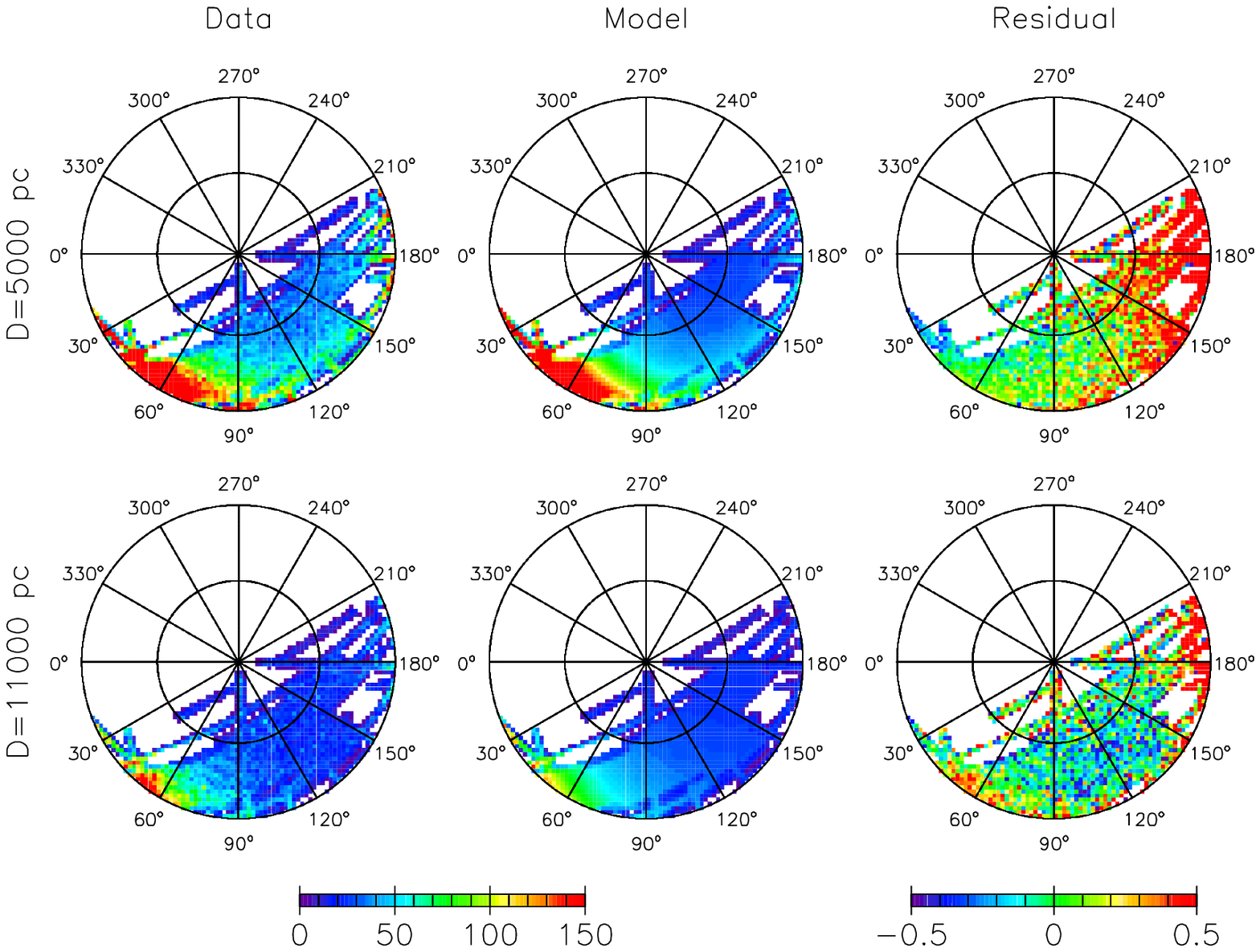}
\vspace{0.5cm}
\caption{Same as Figure~\ref{denn} on southern Galactic hemisphere. Apart from the enhancement at low latitudes in the anticenter region that may be related to the Monoceros stream, the Sagittarius stream
($90^\circ<l<230^\circ$, $b>-40^\circ$), and the Hercules-Aquilae cloud
($l\sim60^\circ$, $b\sim-35^\circ$), no new strong excesses were detected.}
\label{dens}
\end{center}
\end{figure*}

We also present the stellar density maps of the southern Galactic 
hemisphere derived from SDSS DR8 data, in the same distance shells as for the
northern hemisphere (Figure~\ref{dens}).  The SDSS footprint of southern
sky, limited by geographical location of the telescope, is smaller than in
the north, but still large enough to resolve structures on scales as large as
several hundreds deg$^2$.  Apart from the known Hercules-Aquilae
cloud ($l\sim60^\circ$, $b\sim-35^\circ$) and trailing arm of the
Sagittarius dwarf galaxy ($90^\circ<l<230^\circ$, $b>-40^\circ$), there
appear to be no strong overdensities in the halo region. However, we do observe a curious density enhancement in the south at low latitudes in the anticenter direction. This overdensity, that could be associated with the Monoceros stream, or be a feature of the thick disk, will be analyzed in a subsequent publication. Other than the mentioned overdensities, 
the residual maps exhibit a very good fit to the model.

\subsection{Extent of the Virgo Overdensity}

The fact that the VOD is not only visible on the residual plot,
but also as an enhancement on the density plot itself, motivates us for another, model free
look at the data.  This approach also reduces the dependence of our results and conclusions on the details of the analytic Galactic model.

\begin{figure}
\begin{center}
\includegraphics[scale=0.55]{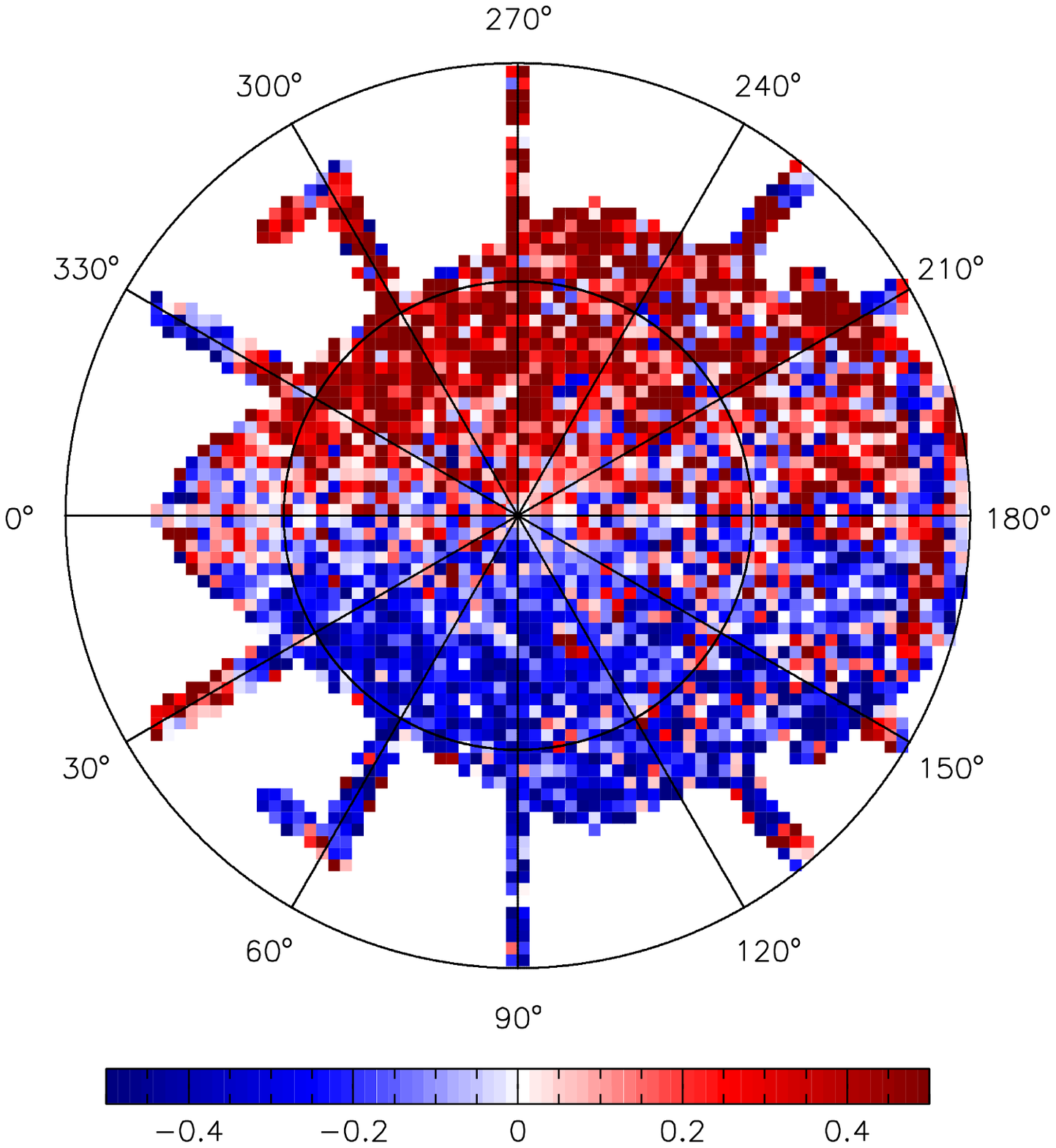}
\caption{A model free confirmation of the Virgo Overdensity is given by a
star count difference between $0^\circ < l < 180^\circ$ and $180^\circ < l <
360^\circ$ regions on the bottom left panel from Figure~\ref{denn}.  The
above region is $\sim50\%$ more densely populated than the lower region. 
The lower limit on the overdensity area is $\sim$2000 deg$^2$, but note that
the overdensity extends down to $b\sim30^\circ$ along the $l=270^\circ$
SEGUE stripe, so the real extent of the Virgo Overdensity might be
considerably larger.}
\label{invert}
\end{center}
\end{figure}

Most overdense pixels on the VOD residual plot (lower right
panel on Figure~\ref{denn}) are above the $l=0^\circ$ line.  If the
halo were symmetric, the halves of the plot divided by this line should also be
symmetric.  In Figure~\ref{invert} we compare the stellar density in these
halves by plotting the value of:
\begin{equation}
f_{W-E}(l,b)=\frac{data(l,b)}{data(360^\circ-l,b)}-1
\label{xinv}
\end{equation}
For a given pixel at Galactic coordinates $(l,b)$, the value of $f_{W-E}$ is simply the
normalized difference between the star counts in the pixel itself
($data(l,b)$) and its counterpart ($data(360^\circ-l,b)$).
The quantity $f_{W-E}$ is defined only for the pixels with axisymmetrical
counterparts, which results in the ``bug like'' shape of the plots.  

The stellar number density was
found to be symmetric on the order of $\lesssim10\%$ at distances closer
than 5 kpc.  However, at 11 kpc the median of $f_{W-E}$ distribution on the upper
half is at $\sim 20\%$. If we adopt the definition of overdensity as density contrast (here defined
with respect to the axisymmetric region) of $\ge$15\%, then the total area of
Virgo overdensity as measured in SDSS DR8 is $\approx2000$~deg$^2$, 
double the size previously measured \citep{J08}.  

Furthermore, we see that the overdensity extends along the $l=300^\circ$ and $270^\circ$ SEGUE stripes.
This indicates that 2000 deg$^2$ is still only a lower limit of its actual size.  We attempt to roughly estimate
the area of the VOD outside the SDSS footprint by
placing an ellipse encompassing most of the pixels with 40\%
overdensity in the SEGUE stripes as well as the overdensity in the main area of the survey.  We find this area to be $\sim$3000
deg$^2$, implying that approximately one third of the VOD may extend
over the area where no data is currently available.

\subsection{Luminosity of the Virgo Overdensity}
To further characterize the
progenitor, we estimate a lower limit on the number of stars associated with
VOD.  The Virgo area was conservatively limited to region subtended by:
($200^\circ<l<210^\circ$, $b>40^\circ$), ($210^\circ<l<270^\circ$,
$b>50^\circ$) and ($270^\circ<l<340^\circ$, $b>60^\circ$).  To estimate the
excess in stellar number counts, we simply subtract the pixel values in
symmetric, Virgo-free area from the VOD pixels, and sum over all the pixels
and distance range (7 - 19 kpc).  This gives the number of VOD stars with
absolute magnitude $M_r=4.0\pm0.5$.  Following \citet{J08}, and assuming the luminosity function of VOD stars is similar to that of the halo (Juri\' c et al., in prep), we obtain an order-of-magnitude estimate of $\sim 10^6$ for the total number of stars present in VOD. This is consistent with numbers found for large globular clusters and dwarf galaxies.

\subsection{Test of the ``Tilted Halo" model}
\begin{figure}
\begin{center}
\includegraphics[scale=0.16]{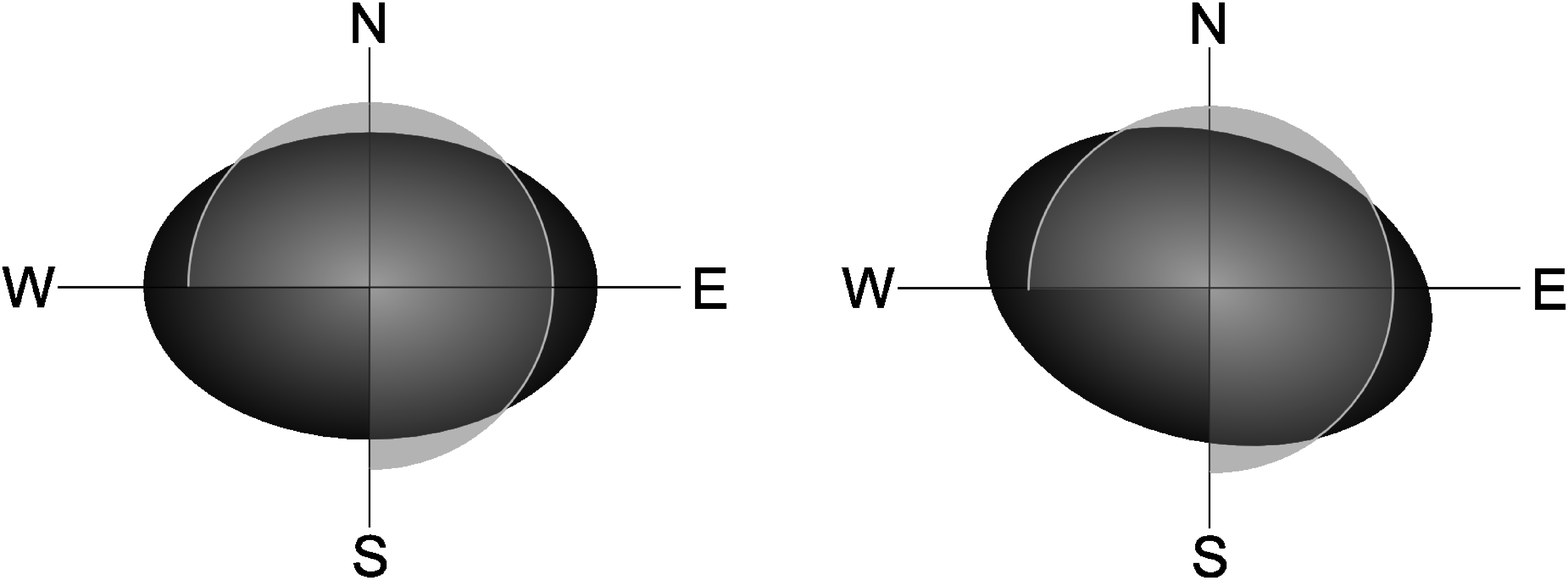}

\caption{Schematic view of the cross section of oblate halo (dark gray) and spherical shell centered on the Sun (light gray). The axis connecting the Sun and the Galactic center is perpendicular to the surface of the plot. In the case of axially symmetric halo (left panel), the stellar density counts at certain distance (as marked by the light gray shell) are symmetric with respect to both $l=0^\circ$ line (W vs. E) and the Galactic plane (N-S). However, if the halo is tilted (right panel), so that we observe excess on the ``Western'' part of the sky (ie the Virgo Overdensity), then the North - South symmetry holds no longer either.}

\label{tilt}
\end{center}
\end{figure}

\begin{figure*}
\begin{center}
\includegraphics[scale=0.7]{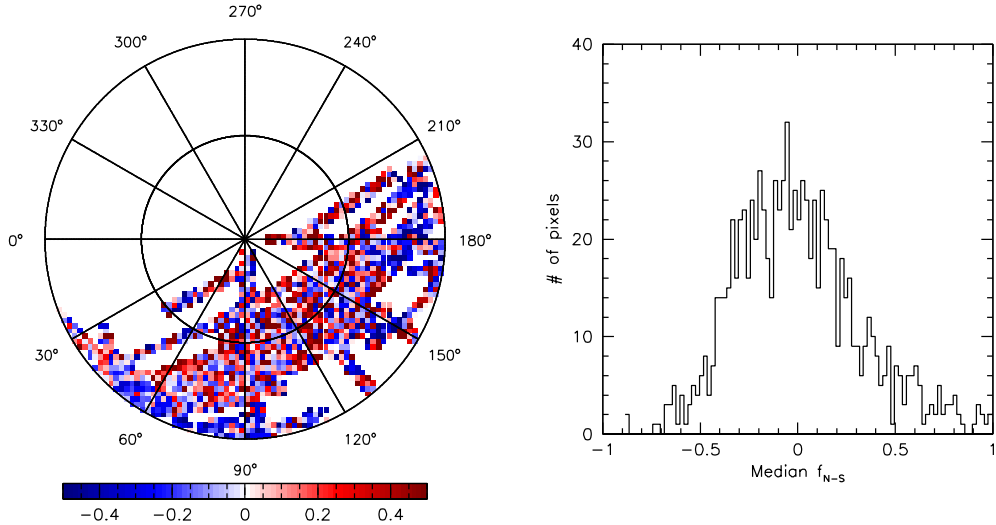}
\caption{A map of difference in star counts in northern and southern Galactic
hemisphere (see bottom left panels on Figures~\ref{denn} and
\ref{dens}, respectively), normalized to the south.  A very good match of the
number counts over 20 kpc range provides strong constraints on the
halo symmetry with respect to the Galactic plane and disfavors the tilted
halo interpretation for the Virgo Overdensity.}
\label{ns}
\end{center}
\end{figure*}

\begin{figure*}
\begin{center}
\includegraphics[scale=0.8]{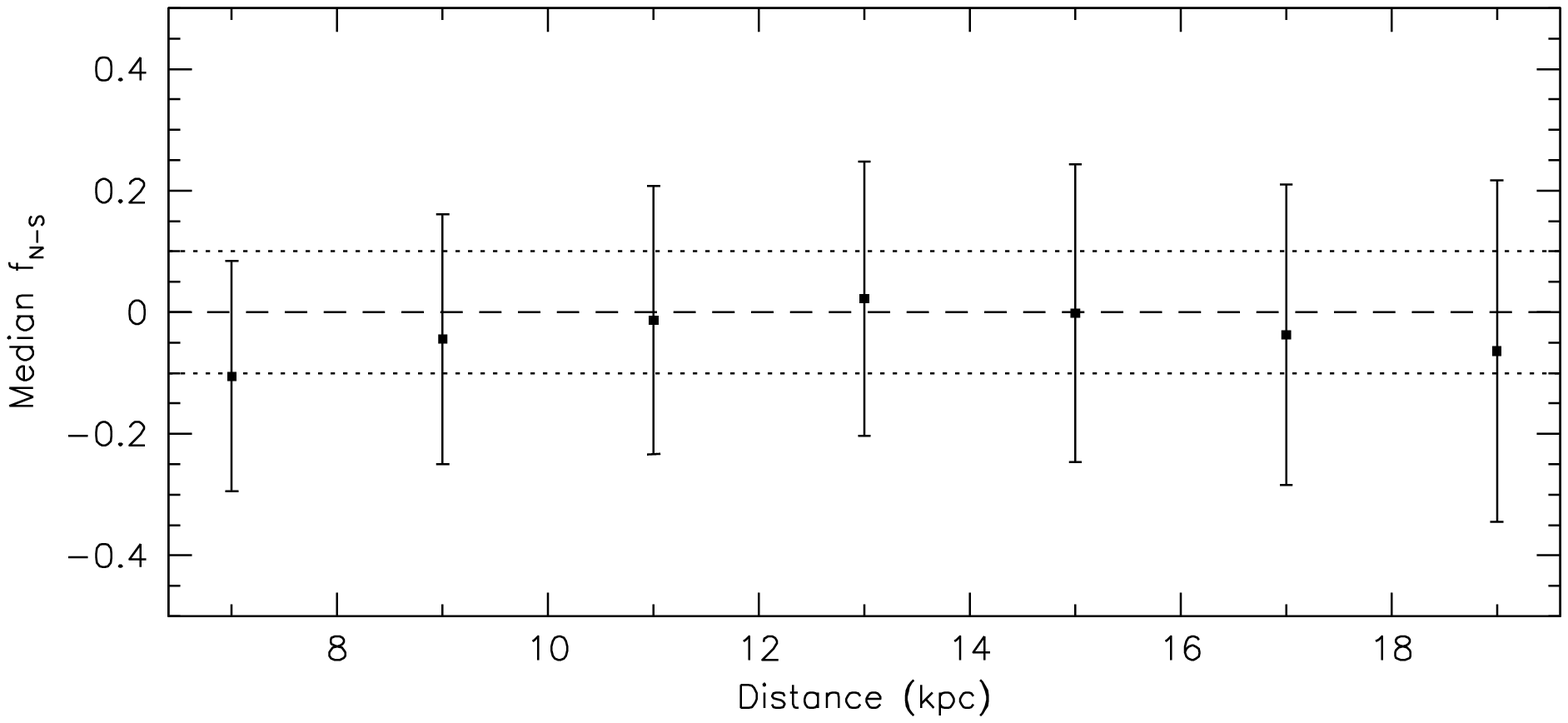}
\caption{Symmetry of stellar halo with respect to the Galactic plane as a
function of distance from the Sun. $f_{N-S}$, as defined in Equation~\ref{yns}, is here computed only for the areas not contaminated by Hercules--Aquilae cloud and Sagittarius stream.
The error bars represent the semi-interquartile
ranges of the $f_{N-S}$ distribution.  The biggest deviation from 
symmetry (denoted by the dashed line at zero) is $\sim10\%$ (dotted line), found in the 7 kpc
distance bin. Note there are no strong asymmetry signatures in the distance range where the VOD is prominent.}
\label{nshisto}
\end{center}
\end{figure*}

Given the large extent of the VOD, an attractive hypothesis is 
that it may not be an overdensity in the stellar halo but a signature of a 
more complex stellar halo density distribution. For example, proposals have been put forward that the Hercules--Aquilae cloud is a signature of dynamical interaction of the disk with the stellar bar, and not a merger event \citep{humphreys2011}.

An oblate halo model, with axes aligned with those of the Galactic disk, does not produce an overdensity signature in maps constructed in Section 2.3. However, if its axes were not aligned with the Galactic disk (eg. if it is ``tilted''), such an overdensity signature will occur (Figure~\ref{tilt}). With the SDSS data
in the southern hemisphere available for a significant portion of the sky, it
is possible to test this tilted halo model by directly comparing the stellar
number density above and below the Galactic plane.
\\

We proceed similarly to steps taken in previous section and plot the values of $f_{N-S}$ defined by:
\begin{equation}
f_{N-S}(l,b)=\frac{data(l,b)}{data(l,-b)}-1
\label{yns}
\end{equation}
in Figure~\ref{ns}. $data(l,b)$ once again denotes stellar number count in
pixel centered at $(l,b)$ and distance of 11 kpc, analagous to the bottom
rows of Figures~\ref{denn} and \ref{dens}.  Note how the value of $f_{N-S}$ is
only defined for those pixels which have data available both above
($b>0$) and below ($b<0$) the Galactic plane.  Although the distribution of
$f_{N-S}$ values is noisy, there is no clear over- or underdensity as would be predicted by Figure~\ref{tilt}.  
This is further confirmed by the distribution of densities in pixels not affected by overdensities 
(excluded regions include Sagittarius dwarf galaxy at $|b|<60^\circ$ and $120^\circ<l<150^\circ$,
$45^\circ<|b|<60^\circ$ and Hercules-Aquilae at circle of radius $7^\circ$
around $l=55^\circ, b=-30^\circ$).  The distribution is slightly asymmetric, but the median is at
-0.01, while the width of the distribution (with semi-interquartile range of
0.22) is consistent with the noise visible on the map of $f_{N-S}$ values.  Given
the distribution centered around zero excess of north versus south, we can
conclude that on the scales of $\sim$20 kpc, halo is symmetric with respect to the
Galactic plane.

This remains true when the exercise is repeated for distance shells in $7-19$ kpc range.  
Figure~\ref{nshisto} shows how the median of $f_{N-S}$
distributions similar to that on Figure~\ref{ns} changes with distance. 
The error bars denote the distributions' IQR.  The plot shows remarkable
consistency in difference of star counts between northern and southern
Galactic hemisphere, the largest deviation being $\sim10\%$ at $7\pm1$ kpc,
and considerably smaller in most other distance bins.

We therefore conclude that a tilted halo cannot account for the existence and observed extent of the Virgo Overdensity.

\section{Discussion}
\label{disc}

Previous studies determined the distance range, radial velocities and
metallicity of stars making up the VOD (see Table~\ref{overview}).  Several authors have also reported evidence of
substructure \citep[e.g.][]{duffau2006}.  In this paper we have used the SDSS DR8 data to reasses the extent of the VOD and find it likely to extend over $\sim3000$ deg$^2$ of the sky, three times more than the original \citet{J08} estimate. 

Stellar overdensities are usually attributed to merger remnants, but
VOD's size and morphology make it unlike a typical tidal stream
thus calling for a more cautious interpretation. An
alternative to the merger remnant hypothesis is that the VOD is a signature of a more complex global halo density distribution (for a
discussion why other proposed hypotheses are less likely, see section 5.5 in
\citealt{J08}). In this paper, we have tested whether a ``tilted'' halo can explain the observed signature.

Figure~\ref{tilt} shows schematically how a tilted halo would appear  as an overdensity in Figures~\ref{invert}~and~\ref{ns}. 
Axisymmetric halo on the left panel would be observed as being completely
symmetric in both $f_{W-E}$ and $f_{N-S}$ values defined earlier, but if the halo
were tilted as seen in the right panel on Figure~\ref{tilt}, a VOD-like
enhancement in stellar number counts would be observed in the ``western''
side of the $f_{W-E}$ plot.  Halo tilt would also be evident on the $f_{N-S}$ plot
because the number counts on the ``south'' would be larger compared to the
symmetric area on the ``north''.  However, as we can see from Figure~\ref{nshisto}, this is not the
case.  In fact, given the difference in stellar number counts between north
and south areas mapped by the SDSS, we can constrain any north--south halo asymmetry to be less than $10\%$.
As the density contrast of the VOD gets as high as $\sim$50\% (see Figure~\ref{ns}), 
it is clear this simple tilted halo model can only amount to a minor contribution 
to the overall enhancement in density. This leaves the merger hypothesis as the most likely\footnote{While one can always invoke even more complex halo shapes to explain the observations, at some point these become unreasonably contrived.}.

While the VOD's morphology is clearly different from those of other streams (eg. Sagittarius, GD1, etc),
its cloudlike appearance is not entirely unique. Other such overdensities have been observed, eg. 
Hercules-Aquila cloud \citep{belokurov2007}, or Triangulum-Andromeda \citep{rp2004}. Similar structures appear in Galactic halo formation simulations as well. For example, the \citet{johnston2008} simulations modeled to match observed properties of Local Group galaxies
predict several cloud-like morphologies with surface brightness $\approx$32.5 mag
arcsec$^{-2}$ in a Milky Way type galaxy. This is broadly consistent with what we see in Virgo Overdensity; however, we do find traces of VOD as close as 7 kpc from the Galactic center, closer than the simulations would predict. Given the simplicity of the simulations, this may not be a serious issue. 

Furthermore, the accretion origin of the VOD is consistent with kinematic measurements
of an RR Lyrae star assumed to belong to the Virgo Stellar Stream
\citep{dinescu2009}. Based on the radial velocity and proper motion
measurement with a baseline of a century,
\citet{dinescu2009} have determined the orbit of an VSS RR Lyrae, having $r_{peri}=11\pm1$ kpc,
 $r_{apo}=89^{+52}_{-32}$ kpc, and a period of
$T=1.2^{+0.6}_{-0.4}$ Gyr. The measured orbit passes through the VOD, indicating that VOD and VSS may be connected structures with a common origin. Also, the high eccentricity of the orbit is consistent with the observed high eccentricities in simulated merger events that result in cloudlike structures.

Taken together, all these lines of evidence point to the VOD being a signature of a high eccentricity merger event observed at perigalacticon. Future kinematic studies are the best way to further verify this conclusion. 
\\

The authors wish to thank Marla Geha, Dana Casetti, and Nhung Ho for reading the early versions of the manuscript and providing thoughtful comments, and Branimir Sesar for help with accessing the SDSS data. M.J. gratefully acknowledges support by NASA through Hubble Fellowship
grant \#HST-HF-51255.01-A awarded by the Space Telescope Science
Institute, which is operated by the Association of Universities for
Research in Astronomy, Inc., for NASA, under contract NAS 5-26555.

Funding for SDSS-III has been provided by
the Alfred P. Sloan Foundation, the Participating Institutions, the
National Science Foundation, and the USDepartment of Energy.
The SDSS-III Web site is http://www.sdss3.org/.

SDSS-III is managed by the Astrophysical Research
Consortium for the Participating Institutions of the SDSS-III
Collaboration including the University of Arizona, the Brazilian
Participation Group, Brookhaven National Laboratory, University
of Cambridge, University of Florida, the French Participation
Group, the German Participation Group, the Instituto de
Astrofisica de Canarias, the Michigan State/Notre Dame/JINA
Participation Group, Johns Hopkins University, Lawrence
Berkeley National Laboratory, Max Planck Institute for Astrophysics,
New Mexico State University, New York University,
Ohio State University, Pennsylvania State University, University
of Portsmouth, Princeton University, the Spanish Participation
Group, University of Tokyo, University of Utah, Vanderbilt University,
University of Virginia, University of Washington, and
Yale University.

\bibliographystyle{apj}
\bibliography{apj-jour,ms}

\end{document}